\documentclass[onecolumn]{sbrt}
\usepackage[english]{babel}
\usepackage[utf8]{inputenc}
\usepackage{graphicx}
\usepackage{float}
\usepackage{amsmath}
\usepackage{amssymb,color}
\usepackage{algorithm}
\usepackage[noend]{algpseudocode}

\begin{document}

\title{Robust Power Allocation and Linear Precoding for Cell-Free and Multi-Cell Massive MIMO Systems}

\author{Erick F. de Almeida and Rodrigo C. de Lamare
\thanks{Erick F. de Almeida, CETUC, Pontifícia Universidade Católica, Rio de Janeiro-RJ, e-mail: ef.almeida@hotmail.com; Rodrigo C. de Lamare, CETUC, Pontifícia Universidade Católica, Rio de Janeiro-RJ, e-mail: delamare@cetuc.puc-rio.br.}
}

\maketitle



\begin{abstract}
Multi-Cell (MC) systems are present in mobile network operations from the first generation to the fifth generation of wireless networks, and considers the signals of all users to a base station (BS) centered in a cell. Cell-Free (CF) systems works with a large number of distributed antennas serving users at the same time. In this context, Multiple-input multiple-output (MIMO) techniques are used in both topologies and result in performance gains and interference reduction. In order to achieve the benefits mentioned, proper precoder design and power allocation techniques are required in the downlink (DL). In general, DL schemes assume perfect channel state information at the transmitter (CSIT), which is not realistic. This paper studies MC and CF with MIMO systems equipped with linear precoders in the DL and proposes an adaptive algorithm to allocate power in the presence of imperfect CSIT. The proposed robust adaptive power allocation outperforms standard adaptive and uniform power allocation. Simulations also compare the performance of both systems frameworks using minimum mean-square error (MMSE) precoders with robust adaptive power allocation and adaptive power allocation.
\end{abstract}

\section{Introduction}

Broadband wireless cellular networks have passed through many generations over the last 30 years. The fifth generation (5G) is now being commercialized by the Mobile Network Operators (MNOs) since its cellular telecommunications standard specifications were defined, national spectrum auctions were performed worldwide for licensing frequency bands, along with user’s data consumption and data rate demand increases. In order to characterize 5G systems, we can cite higher data rates, lower latency, reduced energy consumption, interference management, massive connectivity, and ultra-reliability \cite{ref1} as big advantages when compared to previous technologies.
\par Multiple-input multiple-output (MIMO) wireless technology was adopted in the fourth generation (4G) and it has been under extensive study ever since. MIMO techniques consist of multiple antennas employed at the transmitter (tx) and/or at the receiver (rx). The main advantage of these systems is that they increase dramatically the overall throughput without the need for additional bandwidth \cite{ref2}. By simultaneous transmission of multiple data streams which share the same time-frequency resources, we have some benefits of MIMO technology that help to achieve such significant performance gains: array gain, spatial diversity gain, spatial multiplexing and interference reduction and avoidance \cite{ref3}.
\par As the number of users tends to tremendously increase, massive MIMO technique was proposed and combined with several other enabling techniques to entitle 5G arrival and it will be present in 6G. In terms of telecommunication architectural setup, multiple antennas are implemented to provide high beamforming and spatially multiplexing gain, thus achieving the high spectral efficiency (SE), energy efficiency, and link reliability. The explosive demand for higher data rates and traffic volume is continuous, thereby wireless communication networks are required to provide better coverage, and uniform user performance over a wide coverage area \cite{ref4}.

\par Multi-cellular co-located antenna system (CAS), where the BS is located at the center of each cell, such as the 5G architecture proposal suffer to satisfy these conditions because of performance degradation at cell-edge and inter-cell interference presence in the system \cite{ref5}. Other techniques also emerge to improve wireless communication and assist in the development of the sixth generation (6G). Distributed antenna systems (DAS) were introduced in \cite{ref6} and their setup can be used to minimize cell-edge interference, providing better coverage, since the BS are distributed in a wide area. In addition, cooperative multipoint joint processing (CoMP) is proposed to reduce inter-cell interference \cite{ref7,did}. Combining the advantages of massive MIMO and the  CoMP and DAS techniques \cite{ref7}, the concept of CF systems has emerged, which is strongly considered for 6G networks.

\par In this work, we study MC and CF with MIMO systems equipped with linear minimum mean-square error (MMSE) precoders in the downlink. {Practical implementations of MIMO systems suffer with imperfect channel state information at the transmitter (CSIT), however traditional design of linear precoders and power allocation techniques assume perfect CSIT. Contributing to solve this issue, we propose a different robust adaptive power allocation (R-APA) algorithm design, to allocate power in the presence of imperfect CSIT. Simulation results illustrate the performance of the proposed R-APA algorithm against existing techniques in MC and CF scenarios, where we clearly see how it outperforms standard adaptive and uniform power allocation.} 

{This paper is structured as follows. Section II describes the system models. Section III presents the linear precoding adopted and the proposed R-APA algorithms. In Section IV, we present and discuss the numerical results, which compare CF and MC networks in terms of Sum Rates and Bit Error Rate (BER). Then, we draw some conclusions over both systems and mention some future works in Section V.}

\section{System Models}
A comprehensive model to efficiently perform transmission over downlink (DL) in a MIMO system consists of proper design of precoders and suitable power allocation, for the correct architectural scheme. In order to exploit channel multipath propagation components (which creates an almost stationary fading pattern in space for each BS) \cite{ref8}, appropriate designs enhance data rate, increase performance and suppress multiuser (MU) interference.

\par For conventional MIMO systems linear precoders and power allocation designs, perfect CSIT is assumed \cite{ref3}. In order to obtain CSI, we lean on two types of spectrum usage techniques for the system: Time division duplex (TDD) or frequency division duplex (FDD). Both methods introduce errors in the estimation procedure, which leads to imperfect CSIT. TDD systems employ training pilots to acquire CSIT whereas FDD systems depend on feedback links \cite{ref9}. Furthermore, in TDD systems and reciprocal channels, the uplink (UL) can be used for channel estimation, {assuming the same channel coherence interval for UL training, UL transmission and DL transmission.}
\par Since real world systems do not meet the perfect CSIT assumption and spectrum techniques (used to obtain CSI) introduce imperfections that can heavily degrade the systems, the design of precoding and power allocation approaches which consider imperfect CSIT is of great significance.

\subsection{Cell-Free}

As a DAS architecture, CF Massive MIMO systems consist of having a large number of individually controllable low-cost low-power single antenna access point (APs) $M$ distributed over a wide area for simultaneously serving a small number of user equipment (UE) $K$ \cite{ref10}. One of the challenges in FDD cell-free massive MIMO is the downlink CSI feedback. Because each AP receives the downlink CSI feedback from all users in conventional massive MIMO systems for designing the precoder, the CSI acquisition and feedback overhead will be huge \cite{ref11}. Based on that, the transmission from the APs to the users (DL) and the transmission from the users to the APs (UL) proceed by TDD operation \cite{ref7}.
\par The APs are connected via backhaul network to a Central Processing Unit (CPU), serving all users at the same time-frequency resource, cooperating with its neighbor. Each AP receives the transmission data, performs power allocation and precoding based on CSI obtained by UL and then, using channel reciprocity, sends the information via DL to the receiver antenna.

\begin{figure}[h!]
\centering
\includegraphics[width=0.3\textwidth]{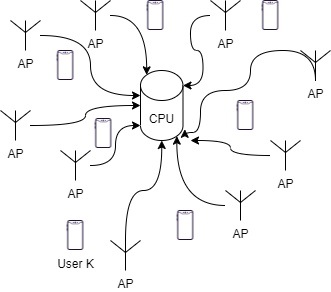}
\caption{Cell-Free massive MIMO System Architecture}
\label{cell-free topology}
\end{figure}

\par There are twofold benefits of the topology in Fig. \ref{cell-free topology}: a reduced data payload in the fronthaul and an increased SE for all UEs \cite{ref12}. However, when compared to CAS, CF has much more backhaul complexity, since the communication between payload data, and power control coefficients is done in the CPU. Assuming $M \gg K$ which means that the user is surrounded by APs, all UEs may have good channel conditions \cite{immse,ref14}.

In MU-MIMO systems, the data are transmitted to \textit{K} users in DL, so the \textit{m}-th AP transmits the symbols $\mathit{x_m}$:
\begin{equation}
\begin{array}{ccl}
x_m = \sqrt{\rho_f} \sum_{k=1}^K p_{mk} a_{mk} s_k,\\
\end{array}
\label{eqn: cf transmitting eq}
\end{equation}

where $\mathit{\rho_f}$ is the transmit power limit of each antenna, $\mathit{p_{mk}}$ is the precoding coefficient and $\mathit{a_{mk}}$ is the power coefficient used by AP \textit{m} for transmission to user \textit{k}, and $\mathit{s_k}$ is the data signal intended for user \textit{k}.

The signal received by the \textit{k}-th user $\mathit{y_k}$ is then represented by:
\begin{equation}
\begin{array}{ccl}
y_k = \sum_{m=1}^M h_{mk}x_m + n_k,\\
\end{array}
\label{eqn: cf received eq}
\end{equation}

where $\mathit{n_k \sim CN(0,\sigma_w^2)}$ and $\mathit{h_{mk}}$ is the estimated channel coefficient, {also complex normal distribution}.

\par For all APs combined we can express the transmitted vector as $\mathbf{x = PNs}$. For all users, we express the received vector:
\begin{equation}
\begin{array}{ccl}
\mathbf{y} = \sqrt{\rho_f} \mathbf{H_{cf}^T x + n},\\
\end{array}
\label{eqn: cf eq}
\end{equation}
{where in this framework, the channel matrix $\mathbf{H_{cf}} \in \mathbb{C}^{{M} \times {K}}$, $\mathbf{P_{cf}} \in \mathbb{C}^{{M} \times {K}}$ is the precoding matrix, $\mathbf{N_{cf}} \in \mathbb{C}^{{K} \times {K}}$ is the power allocation matrix, then $\mathbf{s}$ and $\mathbf{n}  \in \mathbb{C}^{{K}}$ are the symbol and noise vectors, respectively.}

\subsection{Multi-Cell}

{The cellular concept was introduced to increase system's capacity using frequency reuse in a delimited area of coverage. The technique has become essential to expand the number of available channels. To avoid co-channel interference during frequency reuse technique, a cell respects a geographic distance to another cell that uses the same frequency.} 
\par
\begin{figure}[h!]
\centering
\includegraphics[width=0.45\textwidth]{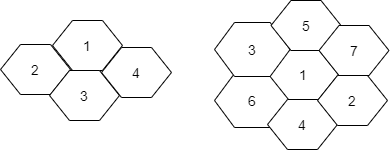}
\caption{Multi-Cell massive MIMO System Architecture (Hexagonal cells cluster: 4 cells on left side and 7 cells on right side)}
\label{cellular system}
\end{figure}

 The base-station (BS) is co-locatted at the center of each cell and a set of cells form a cluster as demonstrated in Fig. \ref{cellular system}. This architecture represents CAS topology for massive MIMO systems. Handover between BS must be considered since the user is served by the BS responsible for that specific cell.

\par Even though this topology reduces channel frequency interference (planned radio access network (RAN) ensures that cells which use the same frequency channel are distant in different clusters), it stills suffer with MUI, inter-cell interference (ICI) and edge users performance reduction.

\par In this setup, the  \textit{k}-th user is equipped with ${N_k}$ antennas and the BS is equipped with ${N_t}$ antennas, where the total number of received antennas is ${N_r}= \sum_{k=1}^{K} {N_k}$. The data are encoded and modulated into a vector of symbols $\mathbf{s}$ $\in$ $\mathbb{C}^{N_r}$. We characterize the precoding matrix that maps the symbols to the transmit antenna as {$\mathbf{P_{mc}} \in \mathbb{C}^{N_t \times N_r}$ and $\mathbf{N_{mc}} \in \mathbb{C}^{N_r \times N_r}$ is the matrix that allocates power to the symbols.}

Once the information is ready for transmission, we can express the transmitted vector $\mathbf{x} \in \mathbb{C}^{N_t}$:

\begin{equation}
\begin{array}{ccl}
\mathbf{x = PNs} = \sum_{k=1}^{N_r} p_k a_k s_k.\\
\end{array}
\label{eqn: system_equation}
\end{equation}

\par Then, for all users the received vector $\mathbf{y} \in \mathbb{C}^{N_r}$:

\begin{equation}
\begin{array}{ccl}
\textbf{y} = \mathbf{H_{mc}x + n},
\end{array}
\label{eqn: mc_equation}
\end{equation}
where in (\ref{eqn: mc_equation})  $\mathbf{H_{mc}} \in \mathbb{C}^{{N_r} \times {N_t}}$ is the multi-cell channel estimation matrix used to transmit the information and $\textbf{n} \in \mathbb{C}^{N_r}$ is the additive noise samples as Gaussian random vector $CN(O,C_{n})$, where $C_{n}$ is the covariance matrix for the noise.

\section{Power Allocation and Precoding}

In this section, we focus on the derivation of the MMSE precoder,  which has better performance ratio between linear precoding option for both systems when compared to ZF in \cite{ref1}. We remark that linear precoders \cite{wence,sint,glcbd,wlbd,rsbd,zf_sec,lrbd} can be replaced by non linear precoders \cite{mbthp,bbprec,rsthp} for enhanced performance. We then present the derivation of the proposed R-APA algorithm that is based on a stochastic gradient (SG) learning strategy \cite{ref15} that takes into account imperfect CSIT. Note that for each system model, namely, MC and CF, the dimensions are different for $\mathbf{P}$, $\mathbf{H}$ and $\mathbf{N}$.

{The ZF precoder can be defined as $\mathbf{P}_{ZF} = \mathbf{H(H^HH})^{-1}$.} 
Considering the $\mathbf{P}_{MMSE}$ corresponds to the following optimization problem:

\begin{equation}
\begin{array}{ccl}
\mathbf{P}_{MMSE} = \arg \min E\{|\mathbf{s} - f\mathbf{y}|{_2^2}\}\\
s. t. ~\mathbb{E} [||\mathbf{x}||_2^2] \leq E_{tx},\\
\end{array}
\end{equation}

{The average  power is obtained through:
\begin{equation}
\begin{split}
\mathbb{E} [|| \mathbf{x}||_2^2] &= \mathbb{E} [(\mathbf{P} \mathbf{N} \mathbf{s})(\mathbf{P} \mathbf{N} \mathbf{s})^H]\\
\end{split}
\end{equation}
}

Then:

\begin{equation}
\begin{array}{ccl}
\mathbf{P}_{MMSE} = \arg \min E\{|\mathbf{s} - f\mathbf{y}|{_2^2}\}\\
s. t. ~ \mathrm{tr(\mathbf{PNC_sN^HP^H})}\leq E_{tx},
\end{array}
\label{eqn: mmse power constraint}
\end{equation}
where ${E_{tx}}$ denotes the total transmit power constraint, and $\mathbf{C_s}$ is the covariance matrix of $s$. Because the transmit
energy is constrained, it is considered that the received signal is scaled with factor $f$ at the receiver, which is part of the optimization.

The MSE cost function is then calculated:
\begin{equation}
\begin{array}{ccl}
\mathbf{C(N)} = E\{|\mathbf{s}-f\mathbf{y}|{_2^2}\}= E[(\mathbf{s} - f\mathbf{y})^H (\mathbf{s} - f\mathbf{y})]
\end{array}
\label{eqn: mse cost function}
\end{equation}
Taking into account (\ref{eqn: cf eq}) and (\ref{eqn: mc_equation}):

\begin{equation}
\begin{array}{ccl}
\mathbf{C(N)} = E[\mathbf{s}^H\mathbf{s}] - fE[\mathbf{s}^H( \sqrt{\rho_f} \mathbf{HPNs} + \mathbf{n}] \\
-fE[(\sqrt{\rho_f}\mathbf{HPNs} + \mathbf{n})^H \mathbf{s}] \\
+ f^2E[(\sqrt{\rho_f}\mathbf{HPNs} + \mathbf{n})^H (\sqrt{\rho_f}\mathbf{HPNs} + \mathbf{n})]
\label{eqn: mse resolution}
\end{array}
\end{equation}

Since s and n are statistically independent: $E[\mathbf{s}^H\mathbf{n}]=E[\mathbf{sn}^H] = 0$; $E[\mathbf{s}^H\mathbf{s}]= \mathbf{Cs}$ and $E[\mathbf{n}^H\mathbf{n}]= \mathbf{Cn}$:

\begin{equation}
\begin{array}{ccl}
\mathbf{C(N)} = tr(\mathbf{Cs}) - ftr(\sqrt{\rho_f}\mathbf{HPNCs})\\
- ftr(\sqrt{\rho_f}\mathbf{N}^H\mathbf{P}^\mathbf{H}\mathbf{H}^H\mathbf{Cs}) + f^2tr(\rho_f\mathbf{N}^H\mathbf{P}^H\mathbf{H}^H\mathbf{HPNCs})\\
+ f^2tr(\mathbf{Cn})
\end{array}
\label{eqn: mse resolution 2}
\end{equation}

After applying the method of Lagrange multipliers, Wirtinger's calculus, equating the terms to zero, taking the derivative with respect to $f$, and using the trace operator, we obtain the following result for the MMSE precoder:
\begin{equation}
\begin{array}{ccl}
\mathbf{P}_{MMSE} = \frac{f}{\sqrt{\rho_f}} (\mathbf{H H}^H + \frac{tr(\mathbf{C_n})}{E_{tx}})^{-1} \mathbf{H}^H \mathbf{N}^{-1},
\end{array}
\label{eqn: mmse precoder}
\end{equation}

\subsection{Adaptative Power Allocation (APA)}

We define the Algorithm~\ref{alg:algorithm_sg_adaptive}, the adaptive SG learning strategy that performs APA as explained in details:

\begin{algorithm}[H]
\caption{APA Algorithm Based on the Stochastic Gradient}
\begin{algorithmic}[1]
\State \textbf{Parameters:} $\mu$ (step size) and $T_{\text{APA}}$ (number of iterations).
\State \textbf{Initialization:} $\eta_{k}[0] = 10^{-3}, k=1,\dots,K.$
\State \textbf{For} i= 0:$T_{\text{APA}}$
\State \quad Set $\mathbf{N}[i]$ as a diagonal matrix with $\sqrt{\boldsymbol{\eta}[i]}$ on its diagonal.
\State \quad Define $\mathcal{C}(\mathbf{N})$.
\State \quad Compute  $\nabla C(N)$.
\State \quad Calculate $\mathbf{N}[i+1] = \mathbf{N}[i] - \mu \hat{\underset{\mathbf{N}^{*}}{\nabla}} \mathcal{C}(\mathbf{N})$
\State \quad Obtain $\mathbf{N}[i+1] = \mathbf{N}[i+1] \odot \mathbf{I}_K$
\State \quad Apply the per-antenna constraint $\boldsymbol{\delta}_{m} \cdot \boldsymbol{\eta}[i+1]  \leq 1, m=1,\dots,M$ to adjust $\mathbf{N}[i+1]$
\State \textbf{end}
\State Obtain $\mathbf{N} = \mathbf{N}[i+1]$.
\end{algorithmic}
\label{alg:algorithm_sg_adaptive}
\end{algorithm}

Based on the algorithm steps, let us consider (\ref{eqn: mse resolution 2}) to calculate the instantaneous gradient, using Wirting's calculus with respect to $\mathbf{N}^*$ and then find the matrix $\mathbf{N}$:

\begin{equation}
\begin{array}{ccl}
\nabla \mathbf{C(N)} = \frac{\partial\mathbf{ C(N)}}{\partial \mathbf{N}^*}\\
\\
\nabla \mathbf{C(N)} = - f\sqrt{\rho_f} \mathbf{P}^H \mathbf{H}^H
\mathbf{Cs} + f^2\rho_f \mathbf{P}^H \mathbf{H}^H \mathbf{HPNCs}
\end{array}
\label{eqn: APA}
\end{equation}

The matrix $\mathbf{N}$ from (\ref{eqn: system_equation}) is updated by:

\begin{equation}
\begin{array}{ccl}
\mathbf{N}[i+1] = \mathbf{N}[i] - \mu\nabla \mathbf{C(N)}
\end{array}
\label{eqn: APA}
\end{equation}

\subsection{Robust Adaptive Power Allocation (R-APA)}

For the R-APA approach, we consider imperfect CSIT. {In this case,
the channel $\mathbf{H} = \hat{\mathbf{H}} + \tilde{\mathbf{H}} \in
C^{N_r \times N_t}$ in MC and $\in C^{M \times K}$ in CF.} The
matrix $\mathbf{\hat{H}}$ represents the channel estimate and the
matrix $\mathbf{\tilde{H}}$ models the CSIT imperfection,
{obtained by considering the error of the estimation procedure} 
Each coefficient $h_{ij}$ of the matrix $\mathbf{H}$ represents the
link between the \textit{i}-th receive antenna and the \textit{j}-th
transmit antenna. The channel matrix can be expressed by $\mathbf{H
= [H_1, H_2, \hdots , H_k]}$, where $\mathbf{H_k}$ denotes the
channel connecting the BS to the kth user.
\par Let us  reconsider the MSE function (\ref{eqn: mse cost function}) and rewrite with the new definition of $\mathbf{H}$:
\begin{equation}
{C({\mathbf{N}_{robust})}= E\{| \mathbf{s} - \mathbf{\hat{H}PNs} -
\mathbf{\tilde{H}PNs} - \mathbf{n}|_2^2} \} \label{eqn: mse
costfunction - r_apa}
\end{equation}

Following the Algorithm 1 steps, let us expand the terms to find $\mathbf{C(N_{robust})}$:

\begin{equation}
\begin{array}{ccl}
\mathbf{C(N_{rob})} = E[(\mathbf{s}^H\mathbf{s}) - (\mathbf{s}^H
\mathbf{\tilde{H}PNs}) - (\mathbf{s}^H\mathbf{\hat{H}PNs}) -
(\mathbf{s}^H \mathbf{n})\\ - (\mathbf{s}^H \mathbf{N}^H
\mathbf{P}^H \mathbf{\hat{H}}^H \mathbf{s}) +
(\mathbf{s}^H \mathbf{N}^H \mathbf{P}^H \mathbf{\hat{H}}^H \mathbf{\hat{H}PNs})\\
+ (\mathbf{s}^H\mathbf{N}^H \mathbf{HP}^H \mathbf{\hat{H}}^H
\mathbf{\tilde{H}PNs}) + (\mathbf{s}^H \mathbf{N}^H \mathbf{P}^H
\mathbf{\hat{H}}^H \mathbf{n}) \\ - (\mathbf{s}^H \mathbf{N}^H
\mathbf{P}^H \mathbf{\tilde{H}}^H \mathbf{s}) + (\mathbf{s}^H
\mathbf{N}^H \mathbf{P}^H \mathbf{\tilde{H}}^H \mathbf{\hat{H} P N
s}) \\ + (\mathbf{s}^H \mathbf{N}^H \mathbf{P}^H
\mathbf{\tilde{H}}^H \mathbf{\tilde{H} P N s}) + (\mathbf{s}^H
\mathbf{N}^H \mathbf{P}^H \mathbf{\tilde{H}}^H \mathbf{n}) \\ -
(\mathbf{n}^H \mathbf{s}) + (\mathbf{n}^H \mathbf{\hat{H} P N s}) +
(\mathbf{n}^H \mathbf{\tilde{H} P N s}) + (\mathbf{n}^H \mathbf{n})]
\end{array}
\label{eqn: mse cost function - r_apa2}
\end{equation}
where $\mathbf{x}$ and $\mathbf{n}$ are statistically independent,
as well as $\mathbf{\tilde{H}}$ and $\mathbf{\hat{H}}$. Applying
trace operator, we have:

\begin{equation}
\mathbf{C(N_{rob})} = tr(\mathbf{C_s}) - 2tr(\mathbf{\hat{H}PNC_s}) \\ +
2tr(\mathbf{N}^H \mathbf{P}^H \mathbf{\hat{H}}^H \mathbf{\hat{H} P N C_s})  + tr(\mathbf{N}^H \mathbf{P}^H \mathbf{\tilde{H}}^H \mathbf{\tilde{H} P N C_s}) \\+ tr(\mathbf{C_n})
\label{eqn: mse cost function - r_apa3}
\end{equation}

The independent vectors $\hat{h}_k$ and $\hat{h}_j$, with $k \neq j$
\begin{equation}
\begin{array}{ccl}
E[\mathbf{\tilde{H}}^H \mathbf{\tilde{H}}]= [\mathbf{\tilde{G}}] = \begin{bmatrix}
\sigma^2_{e_1} & 0 & \hdots & 0\\
0 & \sigma^2_{e_2} & \hdots & 0\\
\vdots & \vdots & \ddots & \vdots\\
0 & 0 & \hdots & \sigma^2_{e_{N_t}}\\
\end{bmatrix}
\end{array}
\label{eqn: imperfect CSI}
\end{equation}

Without loss of generality, we consider $\sigma^2_{e_i}$ = $\sigma^2_{e_j}$, furthermore $\sigma^2_{e_i}$ = $N_r \sigma^2_{e}$
\begin{equation}
\begin{array}{ccl}
[\mathbf{\tilde{G}}] = N_r\begin{bmatrix}
\sigma^2_{e} & 0 & \hdots & 0\\
0 & \sigma^2_{e} & \hdots & 0\\
\vdots & \vdots & \ddots & \vdots\\
0 & 0 & \hdots & \sigma^2_{e}\\
\end{bmatrix}
\end{array}
\label{eqn: imperfect CSI}
\end{equation}

We then derive $\nabla \mathbf{C(N_{rob})}$:

\begin{equation}
\begin{array}{ccl}
\nabla \mathbf{C(N_{rob})} = \frac{\partial \mathbf{C(N_{rob})}}{\partial \mathbf{N}}\\
\\
\nabla \mathbf{C(N_{rob})} = - 2(\mathbf{\hat{H}PC_s})  + 2(\mathbf{P}^H \mathbf{\hat{H}}^H \mathbf{\hat{H} P N C_s}) \\
+ (\mathbf{P}^H [\mathbf{\tilde{G}}] \mathbf{P N C_s})
\end{array}
\label{eqn: recurssion R-APA}
\end{equation}

The result from (\ref{eqn: recurssion R-APA}) can be used to update the matrix $N$ by gradient ascendent recursion:

\begin{equation}
\begin{array}{ccl}
\mathbf{N}[i+1] = \mathbf{N}[i] - \mu\nabla \mathbf{C(N_{rob}})
\end{array}
\label{eqn: APA}
\end{equation}

Including imperfect CSIT to the recursion of power allocation coefficients, increases robustness against CSIT uncertainties. The detection procedure can employ several strategies that differ in the way the interference is mitigated \cite{spa,mfsic,mbdf,bfidd,1bitidd,lrcc,dynovs,listmtc,dynmtc1,dynmtc2}.

\section{Numerical Results}

In  this section, we study the performance of both systems in terms of BER vs. SNR and  SUM-Rate vs. SNR. {Simulations were performed considering QPSK modulation and large-scaling fading models. The dimensions of the systems are represented in each chart result.} 
Three linear precoding techniques, Matched-Filter (MF) or Conjugated Beamforming (CB), Zero-Forcing (ZF) and Minimum Mean Squared Error (MMSE) were used. We also compare Uniform Power Allocation (UPA) and Adaptive Power Allocation (APA) to the R-APA proposed in this work.

\par
\begin{figure}[H]
\centering
\includegraphics[width=0.4\textwidth]{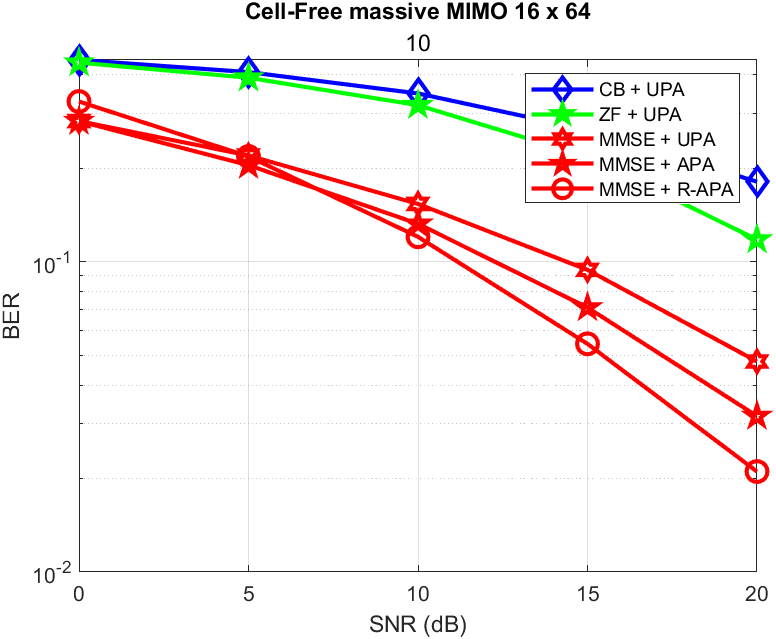}
\caption{Cell-Free BER vs. SNR: K = 16 and M = 64}
\label{cell-free ber vs snr}
\end{figure}

\begin{figure}[H]
\centering
\includegraphics[width=0.4\textwidth]{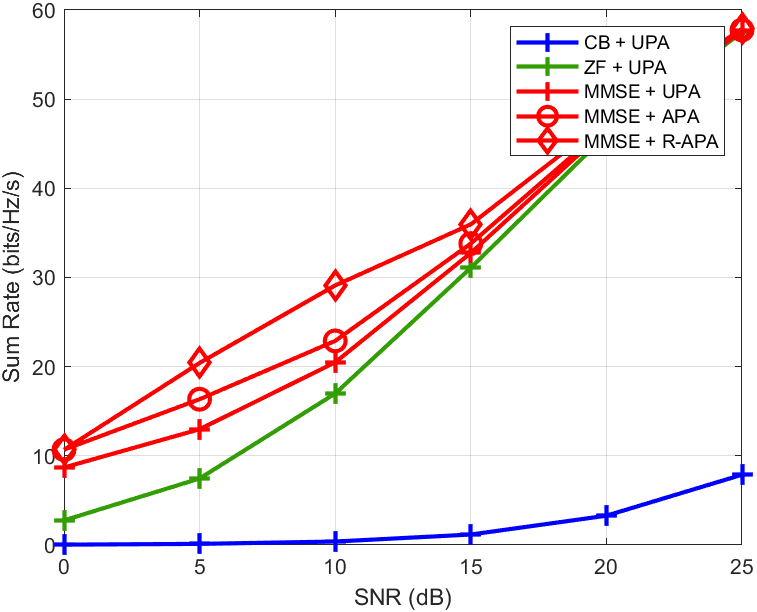}
\caption{Cell-Free Sum-Rate vs. SNR: K = 16 and M = 64}
\label{cell-free sum-rate}
\end{figure}

 We would like to start the analysis with the comparison over the CF performance. We note that for linear precoding schemes, MMSE precoder is the one with better rate in {Fig. \ref{cell-free ber vs snr}} 
Adding to that, APA approach brings a good enhance of performance over UPA. Using CSIT imperfection, we can get even a better improvement. Next, in Fig.  \ref{cell-free sum-rate}  we see the same performance in terms of Sum-Rate vs. SNR, where MMSE precoder with UPA achieves better results than other precoders. R-APA approach increases the outcome even more.

\par
\begin{figure}[H]
\centering
\includegraphics[width=0.4\textwidth]{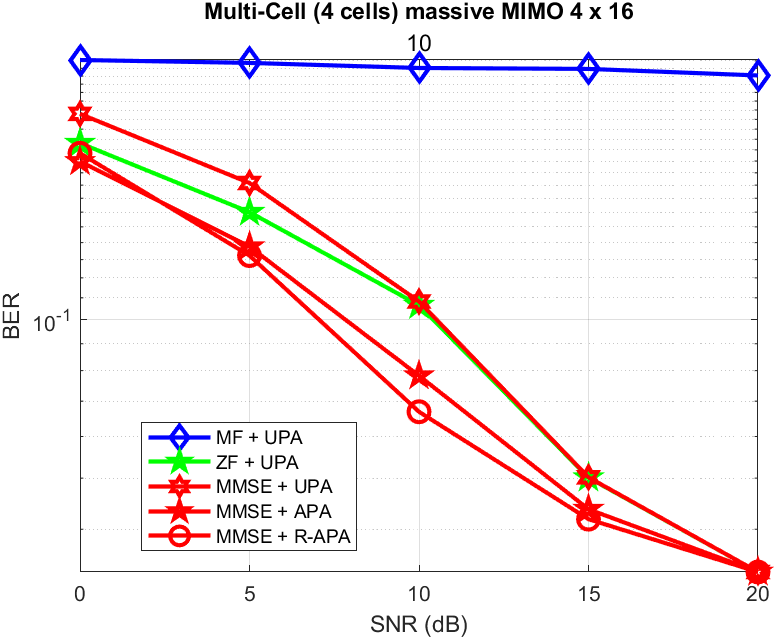}
\caption{Multi-Cell (4 cells) BER vs. SNR: ${N_r}$ = 4 and ${N_t}$ = 16}
\label{mc ber vs snr}
\end{figure}

 The MC architecture is presented in the following simulation.  In { Fig. \ref{mc ber vs snr}} 
we compare, in a four cells system, three types of linear precoders and three power allocation techniques, just as before for DAS. For this scheme, we have the addition of ICI. We can see that MMSE outperfoms ZF and MF (which is almost constant) using UPA. Furthermore, MMSE using APA and R-APA have better results than MMSE + UPA.

\begin{figure}[H]
\centering
\includegraphics[width=0.4\textwidth]{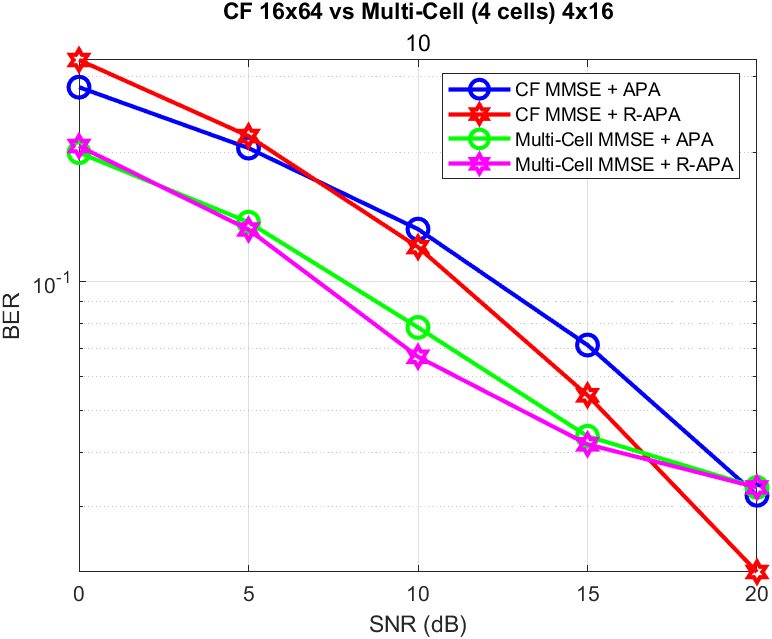}
\caption{BER vs SNR performance: Multi-Cell vs Cell-Free}
\label{cell-free vs mc}
\end{figure}

Finally in { Fig. \ref{cell-free vs mc}} 
, we compare both systems best precoding and power allocation performances. Note that MC outperfoms CF for small values of SNR, however when the curves get close to 15 dB, CF MMSE + R-APA achieves better performance.

\section{Conclusion}
In this work we have presented some massive MIMO concepts, applications, benefit and compared two architecture CAS (low backhaul requirement and all users are served by only one centralized antenna) and DAS (high backhaul requirement and all users are served by all the distributed antennas simultaneously in the coverage area) in terms of BER vs. SNR and Sum-rate vs. SNR.

\par We structured each system architecture for the data transmission and reception. We considered downlink linear precoding, where MMSE precoder was derived and compared to ZF and MF linear approaches. We also calculated the power allocation for an adaptive algorithm that use recursive expressions to recalculate the power allocation matrix taking into account the MSE cost function of the system. Next, we included CSIT imperfections into the power allocation algorithm making it more robust and increasing its rates. Numerical results demonstrate the performance of all precoding designs and power allocation algorithms.

\par {Finally, MIMO MC systems used in nowadays wireless communication networks (4G and 5G) have proven their efficiency, achieving high rates and reasonable system capacity. MIMO CF systems have shown  substantial potential for future wireless technologies, achieving comparable rates to MC systems and excellent system capacity levels. As future works, we can suggest the study of CF high mobility systems, since current systems are considered for scenarios where the mobility of the users are less than 10km/h, due to the application of the TDD protocol and the coherence time interval \cite{ref13}.} 

\end{document}